\documentclass[aps,prmaterials,reprint,superscriptaddress]{revtex4-2}

\usepackage[utf8]{inputenc}
\usepackage[T1]{fontenc}
\usepackage{graphicx}
\usepackage{amsmath}
\usepackage{bm}
\usepackage{multirow}
\usepackage{xcolor}
\usepackage[normalem]{ulem}
\usepackage[makeroom]{cancel}
\usepackage{gensymb}

\begin{document}
\title{Dipolar interlayer excitons in transition metal dichalcogenide alloy heterobilayers}

\author{E.~Katsipoulaki}
\affiliation{Institute of Electronic Structure and Laser, Foundation for Research and Technology-Hellas, Heraklion 71110, Greece}

\author{N.G.~Chatzarakis}
\affiliation{Microelectronics Research Group, IESL-FORTH, Heraklion 70013, Greece}

\author{E.~Rigoutsou}
\affiliation{Department of Materials Science and Engineering, University of Crete, Heraklion 70013, Greece}

\author{D.~Katrisioti}
\affiliation{Institute of Electronic Structure and Laser, Foundation for Research and Technology-Hellas, Heraklion 71110, Greece}
\affiliation{Department of Materials Science and Engineering, University of Crete, Heraklion 70013, Greece}

\author{T.~Taniguchi}
\affiliation{Research Center for Materials Nanoarchitectonics, National Institute for Materials Science, 1-1 Namiki, Tsukuba 305-0044, Japan}

\author{K.~Watanabe}
\affiliation{Research Center for Electronic and Optical Materials, National Institute for Materials Science, 1-1 Namiki, Tsukuba 305-0044, Japan}

\author{S.~Psilodimitrakopoulos}
\affiliation{Institute of Electronic Structure and Laser, Foundation for Research and Technology-Hellas, Heraklion 71110, Greece}

\author{N.T.~Pelekanos}
\affiliation{Microelectronics Research Group, IESL-FORTH, Heraklion 70013, Greece}
\affiliation{Department of Materials Science and Engineering, University of Crete, Heraklion 70013, Greece}

\author{I.~Paradisanos}
\email{iparad@iesl.forth.gr}
\affiliation{Institute of Electronic Structure and Laser, Foundation for Research and Technology-Hellas, Heraklion 71110, Greece}
\affiliation{Department of Materials Science and Engineering, University of Crete, Heraklion 70013, Greece}

\begin{abstract}
Interlayer excitons in transition metal dichalcogenide (TMD) heterobilayers possess a permanent electric dipole moment and long recombination lifetimes, making them a promising platform for exploring excitonic many-body physics. Here, we report dipolar interlayer excitons in a MoS$_{1.4}$Se$_{0.6}$/MoSe$_2$ heterobilayer encapsulated in hexagonal boron nitride. Low-temperature photoluminescence measurements reveal a distinct emission peak at $\sim1.4$~eV, attributed to radiative recombination of interlayer excitons. The emission exhibits a blueshift with increasing excitation power, indicating repulsive dipole--dipole interactions. Time-resolved photoluminescence measurements uncover nanosecond-scale lifetimes, consistent with the spatial separation of electrons and holes across the two layers. These findings establish chalcogen-alloyed TMD heterobilayers as a versatile platform for engineering dipolar excitons and tuning excitonic interactions in van der Waals materials.
\end{abstract}


\maketitle

\section{Introduction}
Two-dimensional transition metal dichalcogenides (TMDs) and their heterostructures provide a versatile platform for exploring excitonic phenomena in the atomically thin limit~\cite{chernikov2014exciton}. In heterobilayers with type-II band alignment, photoexcited electrons and holes can rapidly transfer to different layers, forming spatially indirect interlayer excitons~\cite{fang2014strong, ren2025charge}. Owing to the reduced electron–hole wavefunction overlap, these excitons exhibit long recombination lifetimes~\cite{nagler2017interlayer,montblanch2021confinement,ROSATI2025312} and possess a permanent out-of-plane electric dipole moment~\cite{rivera2015observation}. Such dipolar excitons enable the investigation of many-body interactions~\cite{fogler2014high,wang2019evidence}, exciton transport~\cite{li2021interlayer}, and electrically tunable excitonic devices~\cite{li2020dipolar}. More recently, the field has undergone a rapid evolution with the emergence of moir\'e superlattices in twisted or lattice-mismatched heterobilayers, which impose a periodic potential landscape on interlayer excitons~\cite{tran2019evidence,jin2019observation}. This moir\'e modulation can localize excitons into arrays of quantum emitters with discrete energy levels~\cite{seyler2019signatures,alexeev2019resonantly}, allowing deterministic exciton trapping and the realization of strongly correlated excitonic states. In such systems, the dipolar nature and long lifetime of interlayer excitons enhance exciton–exciton interactions, giving rise to phenomena such as exciton Hubbard physics, ordered dipolar lattices, and signatures of collective quantum phases~\cite{kennes2021moire}. Furthermore, the ability to reach high exciton densities has stimulated intense efforts toward exciton condensation and Bose–Einstein-like phenomena in van der Waals heterostructures~\cite{wang2019evidence}.

\begin{figure*}
\centering
\includegraphics[width=0.85\textwidth]{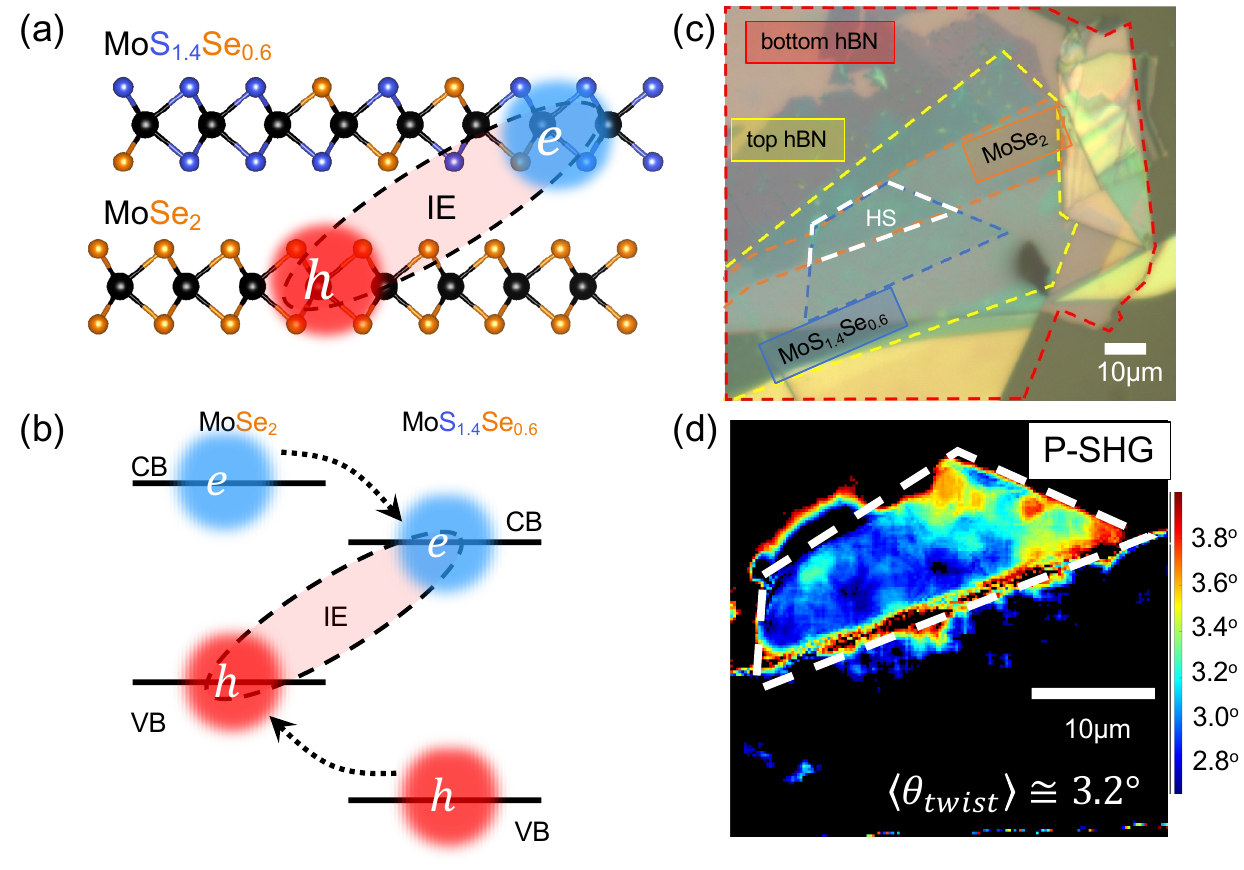}
\caption{
Interlayer exciton formation and crystallographic characterization of the heterobilayer.
(a) Schematic illustration of interlayer excitons (IX) in a MoS$_{1.4}$Se$_{0.6}$/MoSe$_2$ heterobilayer.
(b) Schematic of the type-II band alignment in the MoS$_{1.4}$Se$_{0.6}$/MoSe$_2$ system.
(c) Optical micrograph of the hBN-encapsulated heterobilayer. The white dashed quadrilateral indicates the overlapping region.
(d) Twist-angle map extracted from polarization-resolved second-harmonic generation (P-SHG) measurements, highlighting the heterobilayer region (white dashed quadrilateral).
}
\label{fig1}
\end{figure*}
Beyond stacking different binary TMDs, alloying introduces an additional and still emerging degree of freedom for engineering the electronic band structure and excitonic landscape of two-dimensional materials. TMD alloys allow continuous tuning of band gaps and band offsets~\cite{chen2013tunable,feng2014growth,zhang2014two}, offering new opportunities to tailor interlayer charge transfer processes and exciton energies. In particular, alloying can modify the band alignment, the depth of the moir\'e potential, and the interaction strength of interlayer excitons, thereby allowing controlled tuning of exciton localization and many-body effects. While interlayer excitons have been demonstrated in heterobilayers incorporating transition metal-alloy TMDs compounds~\cite{catanzaro2024resonant,zi2019reversible}, the realization and characterization of dipolar interlayer excitons in heterostructures based on chalcogen-alloyed monolayers remain unexplored.

Here, we present evidence for dipolar interlayer excitons in a MoS$_{1.4}$Se$_{0.6}$/MoSe$_2$ heterobilayer, as schematically illustrated in Fig.~\ref{fig1}(a). The combination of these materials results in a type-II band alignment, in which electrons reside in the MoS$_{1.4}$Se$_{0.6}$ alloy and holes in MoSe$_2$, promoting the formation of dipolar electron--hole pairs (Fig.~\ref{fig1}(b)). Photoluminescence (PL) spectroscopy reveals a distinct low-energy emission at $\sim1.4$ eV, absent in bare monolayers, which we attribute to the radiative recombination of interlayer excitons. A consistent blueshift of the emission energy with increasing excitation power is observed, reflecting dipolar exciton--exciton interactions~\cite{jauregui2019electrical}. Furthermore, time-resolved photoluminescence measurements reveal long-lived exciton dynamics extending to several tens of nanoseconds, supporting their spatially indirect character~\cite{barre2022optical}. These results establish chalcogen-alloyed TMD heterobilayers as a promising platform for engineering dipolar excitons and tailoring excitonic properties in van der Waals heterostructures.

\section{Method}
Bulk crystals of the MoS$_{1.4}$Se$_{0.6}$ alloy were purchased from a commercial supplier (2D Semiconductors) and mechanically exfoliated using standard micromechanical cleavage techniques~\cite{castellanos2014deterministic}. Monolayer flakes were identified by optical contrast and subsequently aligned with respect to MoSe$_2$ monolayers, followed by deterministic dry transfer to form heterobilayers. The final sample consists of a MoS$_{1.4}$Se$_{0.6}$/MoSe$_2$ heterobilayer encapsulated in hexagonal boron nitride (hBN) for optimum optical quality~\cite{cadiz2017excitonic,taniguchi2007synthesis,shree2021guide}. An optical microscopy image of the sample is presented in Fig.~\ref{fig1}(c). The relative crystallographic orientation of the two monolayers was determined using polarization--resolved second--harmonic generation (P-SHG) microscopy~\cite{psilodimitrakopoulos2019twist,psilodimitrakopoulos2021optical}. The measurements yield a mean twist angle of $3.2^{\circ}$ with a standard deviation of $\sigma = 0.34^{\circ}$, indicating a near-aligned heterobilayer configuration. The corresponding P-SHG map of the heterobilayer is shown in Fig.~\ref{fig1}(d). While perfect alignment ($\sim0^{\circ}$) is expected to maximize interlayer coupling and optical efficiency, the present twist angle remains sufficiently small to minimize momentum mismatch at the band edges.

\begin{figure*}
\centering
\includegraphics[width=\textwidth]{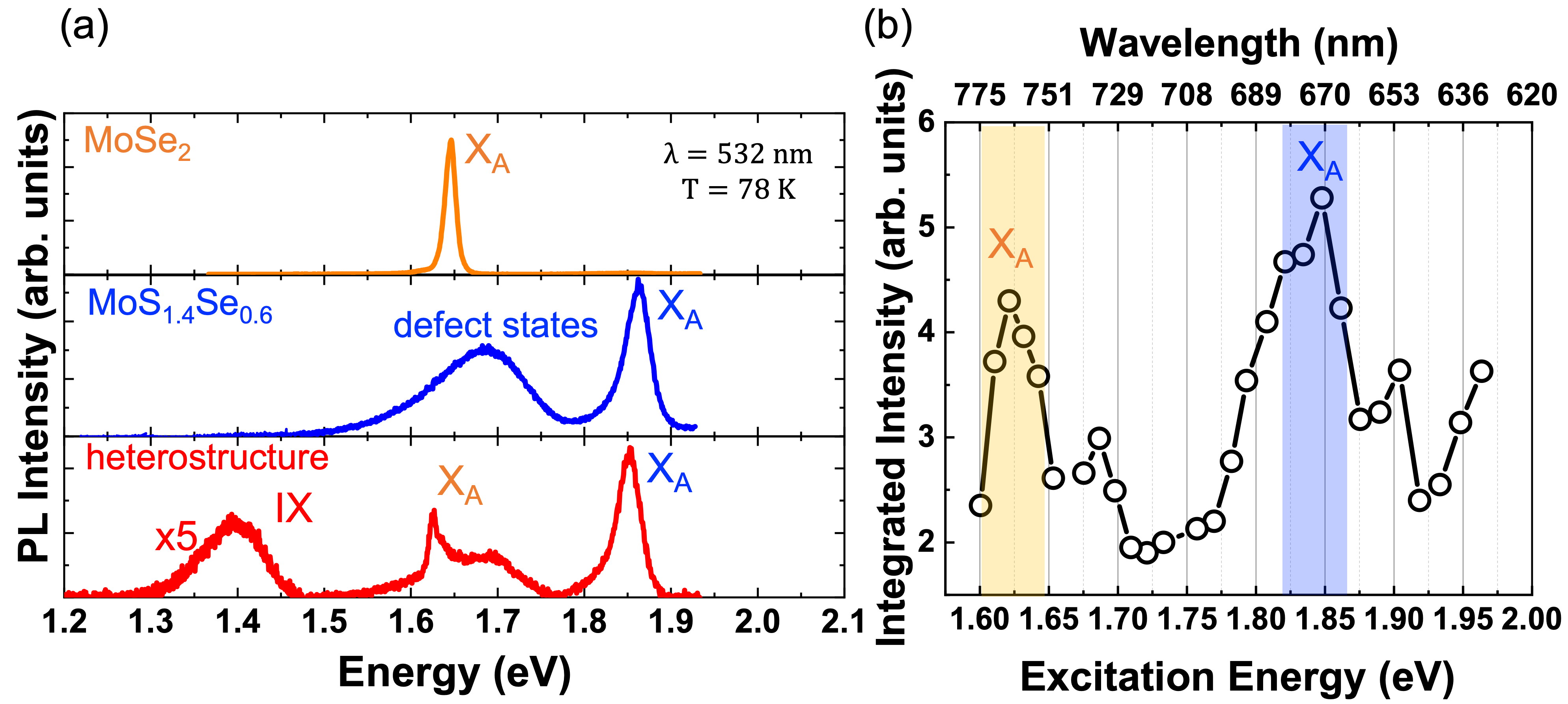}
\caption{
PL spectroscopy of the individual monolayers and the heterobilayer.
(a) Low-temperature ($T=78$~K) PL spectra of the MoSe$_2$ monolayer (orange), the MoS$_{1.4}$Se$_{0.6}$ alloy monolayer (blue), and the MoS$_{1.4}$Se$_{0.6}$/MoSe$_2$ heterobilayer (red) under 532~nm excitation ($P=20\,\mu$W). The spectra display the intralayer excitonic transitions ($X_A$ and $X_B$) of the individual layers, along with defect-related emission. The heterobilayer spectrum additionally exhibits a low-energy interlayer exciton (IX) emission (multiplied by a factor of 5 for clarity).
(b) Photoluminescence excitation (PLE) spectroscopy of the heterobilayer, showing the integrated intensity of the IX emission as a function of excitation energy.
}
\label{fig2}
\end{figure*}
\section{Results and Discussion}
PL spectroscopy is employed to probe the excitonic properties of the individual monolayers and the heterobilayer. Fig.~\ref{fig2}(a) shows low-temperature PL spectra measured at $T=78$~K under 532~nm excitation ($P=20\,\mu$W) for the MoSe$_2$ monolayer (orange), the MoS$_{1.4}$Se$_{0.6}$ alloy monolayer (blue), and the MoS$_{1.4}$Se$_{0.6}$/MoSe$_2$ heterobilayer (red). The PL spectrum of monolayer MoSe$_2$ is dominated by the intralayer A-exciton ($X_A$) at $\sim1.64$~eV~\cite{ross2013electrical}. The absence of a trion peak indicates a low intrinsic carrier density. For the MoS$_{1.4}$Se$_{0.6}$ alloy monolayer, the spectrum is dominated by $X_A$ at $\sim1.87$~eV. In addition, a broader emission centered at $\sim1.68$~eV is associated with defect-related states. In the heterobilayer, intralayer exciton emission from both constituents is still observed, albeit with reduced intensity and redshifted by $2$--$3$~meV, which we attribute to efficient interlayer charge transfer and enhanced dielectric screening compared to the hBN-encapsulated monolayer regions~\cite{seyler2019signatures}. Notably, the heterobilayer spectrum exhibits the emergence of a new low-energy emission peak at $\sim1.4$~eV, absent in the individual monolayers. To examine the origin of this feature, we perform photoluminescence excitation (PLE) spectroscopy on the heterobilayer, as shown in Fig.~\ref{fig2}(b). The integrated intensity of the $\sim1.4$~eV emission is monitored as a function of excitation energy across the spectral range of the intralayer excitons in both MoSe$_2$ and MoS$_{1.4}$Se$_{0.6}$. A pronounced enhancement of the emission is observed when the excitation energy is resonant with the intralayer exciton transitions of both materials (highlighted by orange and blue shaded regions, respectively). This behavior provides strong evidence that the low-energy emission originates from interlayer excitonic states (IX), formed via charge transfer following intralayer exciton excitation.

\begin{figure*}
\centering
\includegraphics[width=\textwidth]{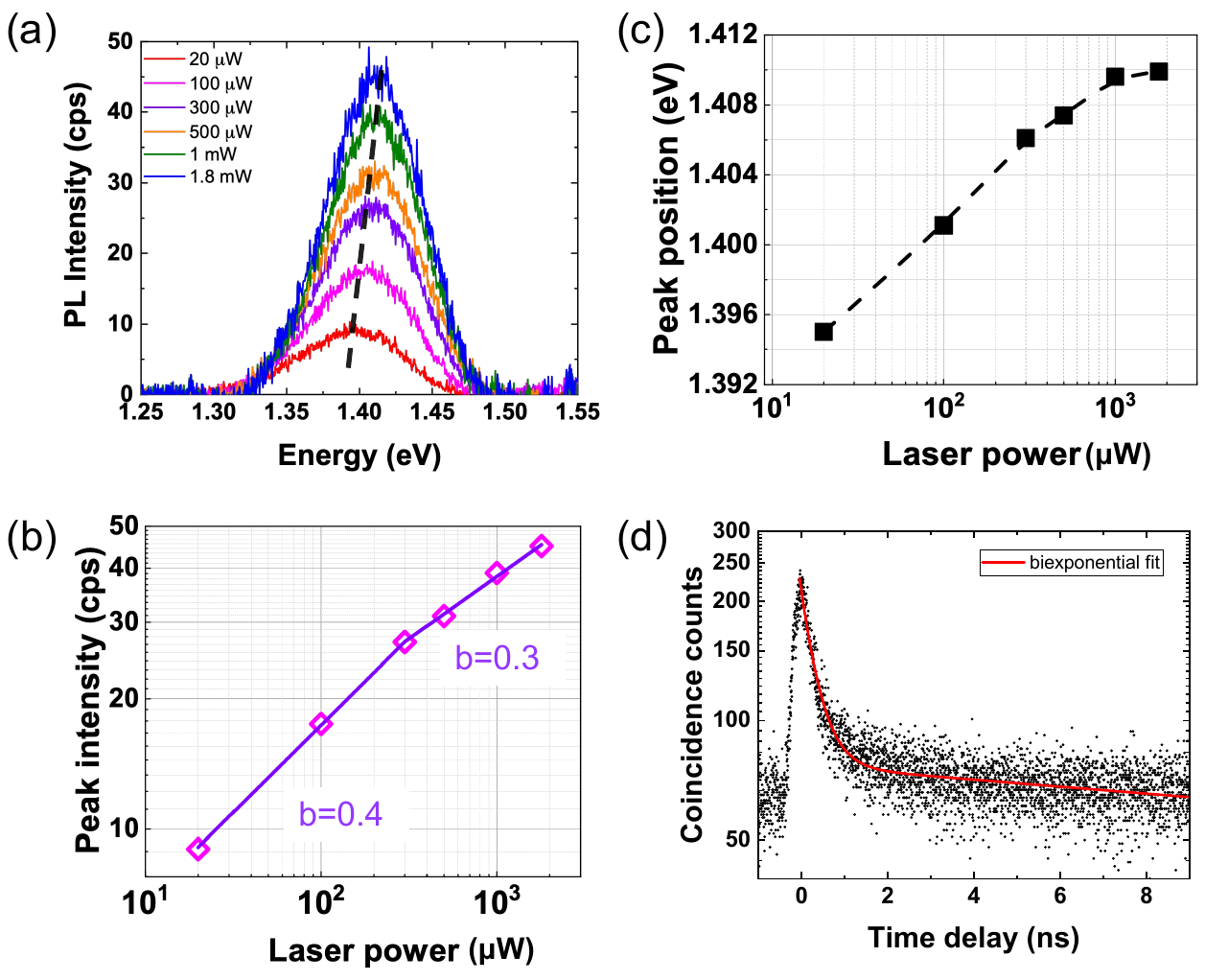}
\caption{
Power dependence and dynamics of interlayer excitons.
(a) PL spectra of the interlayer exciton (IX) emission at $T=78$~K for different excitation powers.
(b) IX intensity as a function of excitation power. The data follow a power-law scaling $I \propto P^{b}$ with exponents $b=0.4$ and $b=0.3$.
(c) Peak energy of the IX emission as a function of excitation power, showing a blueshift with increasing power.
(d) Time-resolved photoluminescence (TRPL) of the IX emission at $T=78$~K, measured using time-correlated single-photon counting. Black dots represent the measured photon counts, while the red curve is a biexponential fit. The extracted decay times consist of a fast component of 374~ps and a long-lived component of 49~ns, consistent with the spatially indirect and dipolar nature of the interlayer excitons.
}
\label{fig3}
\end{figure*}

Further evidence for the dipolar nature of the emission is obtained from excitation power-dependent PL measurements. Fig.~\ref{fig3}(a) shows PL spectra of the interlayer exciton (IX) for excitation powers ranging from $20\,\mu$W to 1.8~mW under 532~nm illumination with a spot size $\sim$1 $\mu$m. The IX intensity as a function of excitation power is presented in Fig.~\ref{fig3}(b). The data are well described by a power-law dependence $I \propto P^{b}$, yielding exponents of $b \approx 0.4$ at low excitation power and $b \approx 0.3$ at higher excitation densities, indicative of pronounced sublinear behavior.
Such sublinear scaling is characteristic of interacting excitonic systems and reflects the presence of nonlinear exciton dynamics at elevated densities, leading to reduced radiative efficiency. In particular, the long lifetime of interlayer excitons promotes processes such as exciton--exciton annihilation~\cite{wietek2024nonlinear}, which can limit the growth of the exciton population with increasing excitation power. In addition, the IX emission exhibits a systematic blueshift with increasing excitation power, as shown in Fig.~\ref{fig3}(c). This behavior arises from repulsive interactions between interlayer excitons, each carrying a permanent out-of-plane dipole moment, which increases the average interaction energy of the exciton ensemble~\cite{steinhoff2024exciton}. Such a density-dependent blueshift is a characteristic signature of dipolar excitons and provides further evidence for the interlayer nature of the observed emission.

Time-resolved photoluminescence (TRPL) measurements provide significant insight into the nature of the IX emission. Fig.~\ref{fig3}(d) shows the temporal evolution of the PL signal following pulsed excitation (wavelength: 375~nm, pulse width: 52~ps). The decay dynamics are well described by a biexponential function, yielding a fast component of $\tau_1 \approx 374$~ps and a slower component of $\tau_2 \approx 49$~ns. We interpret these two timescales as evidence of different relaxation and recombination pathways within the interlayer-exciton manifold. The sub-nanosecond component likely reflects the redistribution of initially populated bright or weakly localized interlayer excitons, including relaxation into momentum-indirect or moir\'e/localized states or exciton--exciton annihilation processes~\cite{wietek2024nonlinear} (as it arises also from Fig.~\ref{fig3}(b)). In contrast, the long-lived component is attributed to recombination from a weakly emissive interlayer-exciton reservoir, whose lifetime is extended due to reduced electron--hole wavefunction overlap and residual momentum mismatch associated with the $\sim3^{\circ}$ twist angle. These results indicate a population transfer from initially emissive interlayer excitons into long-lived semi-dark or localized states, followed by slow radiative or phonon-assisted recombination, consistent with the spatially indirect nature of the interlayer excitons.

\section{Conclusions}
In conclusion, we demonstrate the formation of dipolar interlayer excitons in a MoS$_{1.4}$Se$_{0.6}$/MoSe$_2$ heterobilayer. The emergence of a distinct low-energy emission, together with its resonant enhancement via intralayer exciton excitation, provides clear evidence of efficient interlayer charge transfer and the formation of excitons with spatially indirect character. The observed excitation power-dependent blueshift reveals repulsive dipole--dipole interactions, confirming the dipolar nature of the interlayer excitons. Furthermore, time-resolved measurements uncover long-lived excitonic populations and relaxation pathways involving momentum-dark or localized states, consistent with reduced electron--hole wavefunction overlap and finite momentum mismatch. These results establish chalcogen-alloyed TMD heterostructures as a versatile platform for engineering dipolar excitons, offering tunability of band alignment, exciton interactions, and recombination dynamics. More broadly, alloy heterobilayers provide new opportunities for tailoring excitonic landscapes and exploring many-body phenomena in two-dimensional materials.

\begin{acknowledgments}
S.P. acknowledges support from the EU Horizon-CSA project \textit{DemosAxia} (Grant No.~101160387). K.W. and T.T. acknowledge support from CREST (Grant No.~JPMJCR24A5), JST, and the World Premier International Research Center Initiative (WPI), MEXT, Japan.
\end{acknowledgments}

\bibliographystyle{apsrev4-2}
\bibliography{alloy_HS_biblio}

\end{document}